\documentclass[a4paper,11pt]{article}
\usepackage{pos}

\usepackage{siunitx}

\usepackage{graphicx}
\usepackage{subcaption}
\usepackage{placeins}

\usepackage{hyperref}

\usepackage{braket}

\title{Studies on finite-volume effects in the inclusive semileptonic decays of charmed mesons}
\ShortTitle{Finite-volume effects in inclusive semileptonic decays of charmed mesons}

\author*[a]{Ryan Kellermann}
\author[b,c,d]{Alessandro Barone}
\author[a]{Shoji Hashimoto}
\author[c, d, e]{Andreas Jüttner}
\author[a, ef]{Takashi Kaneko}

\affiliation[a]{Theory Center, Institute of Particle and Nuclear Studies, High Energy Accelerator Research Organization (KEK), Tsukuba 305-0801, Japan and School of High Energy Accelerator Science, The Graduate University for Advanced Studies (SOKENDAI), Tsukuba 305-0801, Japan}

\affiliation[b]{PRISMA+ Cluster of Excellence \& Institut f\"ur Kernphysik, Johannes-Gutenberg-Universit\"at Mainz, D-55099 Mainz, Germany}

\affiliation[c]{School of Physics and Astronomy, University of Southampton, Southampton SO17 1BJ, United Kingdom}

\affiliation[d]{STAG Research Center, University of Southampton, Southampton SO17 1BJ, UK}

\affiliation[e]{CERN, Theoretical Physics Department, Geneva, Switzerland}

\affiliation[f]{Kobayashi-Maskawa Institute for the Origin of Particles and the Universe, Nagoya University, Aichi 464–8602, Japan}

\emailAdd{kelry@post.kek.jp}
\abstract{We report on the calculation of the inclusive semileptonic decay of the $D_s$ meson on the lattice. We simulate the $D_s \rightarrow X_s\ell\nu_\ell$ process with M\"obius domain-wall charm and strange quarks, whose masses were approximately tuned to the physical values. We cover the whole kinematical region. The focus of this work is on the systematic error due to finite-volume effects. We construct a model of two-body final states to describe the data on a finite volume lattice of $L \simeq \SI{0.055}{\femto\meter}$ to investigate the extrapolation to the infinite-volume limit.}

\FullConference{The 40th International Symposium on Lattice Field Theory (Lattice 2023)\\
July 31st - August 4th, 2023\\
Fermi National Accelerator Laboratory\\}

\begin{document}
\maketitle

\section{Introduction}

%In view of the still unresolved discrepancy between exclusive and inclusive determination of the CKM parameters $|V_{ub}|$ and $|V_{cb}|$~\cite{Workman:2022ynf}, we address the calculation of the inclusive rate using lattice correlators proposed in \cite{Gambino:2020crt, Gambino:2022dvu, Hansen:2017mnd, Hansen:2019idp, Bulava:2021fre} employing the Chebyyshev approximation.

%In recent years, experiments have revealed a puzzling tension in B-decays, namely, in the determination of the CKM parameters $|V_{ub}|$ and $|V_{cb}|$ from exclusive and inclusive methods~\cite{Workman:2022ynf}. This discrepancy provides an opportunity for theorists to improve their understanding of these decays. Furthermore, the search for new physics requires precise theoretical predictions from the Standard Model. In view of these points, recently, ideas to extend the application of lattice QCD towards the description of inclusive decays have been proposed~\cite{Gambino:2020crt, Gambino:2022dvu, Hansen:2017mnd, Hansen:2019idp, Bulava:2021fre}. These approaches utilize either the Chebyshev approximation or the Backus-Gilbert approach to obtain the energy integral of the hadronic tensor, which defines the inclusive decay rates. In this paper, we do not address the analysis strategy used to analyze inclusive decays from lattice correlators. We refer to \cite{Gambino:2020crt, Gambino:2022dvu, Kellermann:2022mms} for an overview and \cite{Barone:2023tbl} for a comparison between the Chebyshev and Backus-Gilbert approaches.
We update on our work to calculate the inclusive semileptonic decay rate of the $D_s$-meson. Following our recent work to understand the systematic error associated with the Chebyshev approximation of the kernel function \cite{Kellermann:2022mms}, in this paper we report on our estimate of the systematic error due to finite-volume effects. We refer to \cite{Gambino:2020crt, Gambino:2022dvu, Kellermann:2022mms, Hansen:2017mnd, Hansen:2019idp, Bulava:2021fre} for details on the strategy to calculate inclusive semileptonic decay rates in lattice QCD. In particular, the most recent work \cite{Barone:2023tbl} presents a comparison between the Chebyshev and Backus-Gilbert approaches to approximate the kernel function in the energy integral.

The rest of this paper is structured as follows. We briefly review the inclusive semileptonic decay on the lattice in Sec. \ref{sec:InclusiveLattice}. In Sec. \ref{sec:ModelStrategy} we introduce the model used to estimate finite volume effects before applying it to the lattice data in order to extrapolate towards the infinite volume limit in Sec. \ref{sec:SysErrFinVol}. We present the conclusions in Sec. \ref{sec:Conclusion}.
%The rest of this paper is structured as follows. We highlight how finite-volume effects affect our analysis and present our strategy on how we estimate the associated error in Sec. \ref{sec:ModelStrategy}. %In sec. \ref{sec:Numerics} we discuss the lattice setup used to generate our data. 
%This is followed by a discussion on how the model can be used to fit the lattice data and extrapolate towards the infinite volume limit in Sec. \ref{sec:SysErrFinVol}. We summarize our results and discuss future prospects in Sec. \ref{sec:Conclusion}.

\section{Inclusive semileptonic decays on the lattice}
\label{sec:InclusiveLattice}

The total decay rate of the inclusive semileptonic decay is written as
\begin{align}
    \Gamma \sim \int_0^{\pmb{q}^2_{\text{max}}} d\pmb{q}^2 \sqrt{\pmb{q}^2} \sum_{l=0}^{2} \bar{X}^{(l)}(\pmb{q}^2) \, ,
\end{align}
where $\bar{X}^{(l)}(\pmb{q}^2)$ contains the integral over the hadronic final state energy $\omega$
\begin{align}
    \begin{split}
    \bar{X}_\sigma^{(l)}(\pmb{q}^2) &= \int_{\omega_0}^{\infty} d\omega \, W^{\mu\nu}(\pmb{q},\omega) e^{-2\omega t_0} K^{(l)}_{\mu\nu, \sigma} (\pmb{q},\omega) \\
    &= \braket{\psi^\mu(\pmb{q})|K^{(l)}_{\mu\nu, \sigma}(\pmb{q},\hat{H})|\psi^\nu(\pmb{q})} \, ,
    \end{split}
\end{align}
with the hadronic tensor $W^{\mu\nu}(\pmb{q},\omega)$, $\ket{\psi^{\nu}(\pmb{q})} = e^{-\hat{H} t_0} \tilde{J}^{\nu}(\pmb{q}, 0) \ket{D_s} / \sqrt{2M_{D_s}}$ and $\tilde{J}^{\nu}(\pmb{q}, 0)$ being the Fourier transformed currents. The lower limit  $0 \leq \omega_0 \leq \omega_{\text{min}}$ can be chosen freely as there are no states below the lowest lying energy state $\omega_{\text{min}}$. The parameter $t_0$ is introduced to avoid the contact term which receives contributions from the opposite time ordering corresponding to unphysical states. In the definition of $\bar{X}_\sigma^{(l)}(\pmb{q}^2)$ above 
\begin{align}
   K^{(l)}_{\mu\nu, \sigma}(\pmb{q},\omega) = e^{2\omega t_0} \sqrt{\pmb{q}^2}^{2-l} (m_{D_s} - \omega)^l \theta_{\sigma}(m_{D_s} - \sqrt{\pmb{q}^2} - \omega) \, ,
   \label{equ:KernelFunction}
\end{align}
defines the \textit{kernel function} and $\theta_\sigma(x)$ is a sigmoid function with smearing width $\sigma$.

On the lattice we compute
\begin{align}
    C_{\mu\nu}(t) = \frac{1}{2M_{D_s}} \braket{D_s|\tilde{J}^{\mu\dagger}(\pmb{q},0) e^{-\hat{H}t} \tilde{J}^{\nu}(\pmb{q},0)|D_s} \, ,
    \label{equ:FourPointCorrelator}
\end{align}
and the calculation of the inclusive decay rate is reduced to the one of finding an appropriate polynomial approximation of the kernel function $K^{(l)}_{\mu\nu, \sigma}(\pmb{q},\hat{H})$.

We employ the shifted Chebyshev polynomials $\tilde{T}_j(x)$, with $x = e^{-\omega}$ and define the approximation as
\begin{align}
  \braket{K_{\mu\nu, \sigma}^{(l)}} \simeq \frac{1}{2} \tilde{c}_{0}^{(l)} \braket{\tilde{T}_0} + \sum_{k=1}^{N} \tilde{c}_{k}^{(l)} \braket{\tilde{T}_k} \, .
  \label{equ:ChebApprox}
\end{align}
Here, $\tilde{c}_k^{(l)}$ are analytically known coefficients and $\braket{\tilde{T}_k}$ are referred to as \textit{Chebyshev matrix elements}. We use the notation $\braket{\cdot} \equiv \braket{\psi^\mu|\cdot|\psi^\nu}/\braket{\psi^\mu|\psi^\nu}$. For simplicity, we skip the indices $\mu, \nu$ going forward.

The matrix elements are extracted from a fit to the correlator data following
\begin{align}
    \bar{C}(t) = \sum_{j=0}^{t} \tilde{a}_j^{(t)} \braket{\tilde{T}_j} \, ,
    \label{equ:FitCorrelator}
\end{align}
where $\tilde{a}_j^{(t)}$ are obtained from the power representation of the Chebyshev polynomials, see (A.24) and (A.25) of \cite{Barone:2023tbl} for the definition of $\tilde{a}_j^{(t)}$, and $\bar{C}(t)$ is constructed from \eqref{equ:FourPointCorrelator} as $\bar{C}(t) = C(t+2t_0)/C(2t_0)$. To maximize the available data we choose $t_0 = 1/2$.
We use priors to ensure that the fitted Chebyshev matrix elements satisfy the condition that the Chebyshev polynomials are bounded, i.e. $\left|\braket{\tilde{T}_j}\right| \leq 1$. We refer to \cite{Barone:2023tbl} for more details on the Chebyshev approximation and the practical application.

\section{Modeling strategy}
\label{sec:ModelStrategy}

\begin{figure}
  \centering
  \begin{subfigure}{0.49\textwidth}
    \centering
    \includegraphics[width=0.8\textwidth]{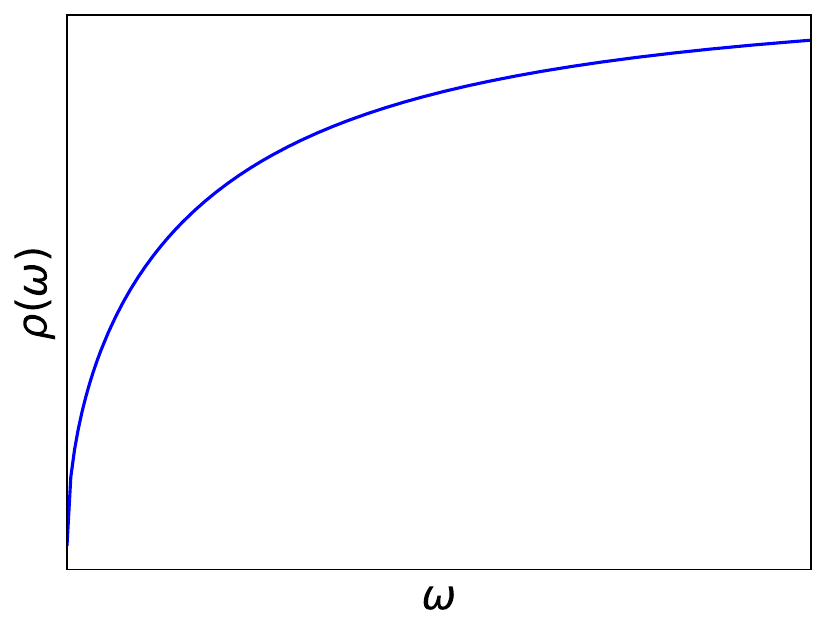}
  \end{subfigure}
  \begin{subfigure}{0.49\textwidth}
    \centering
    \includegraphics[width=0.8\textwidth]{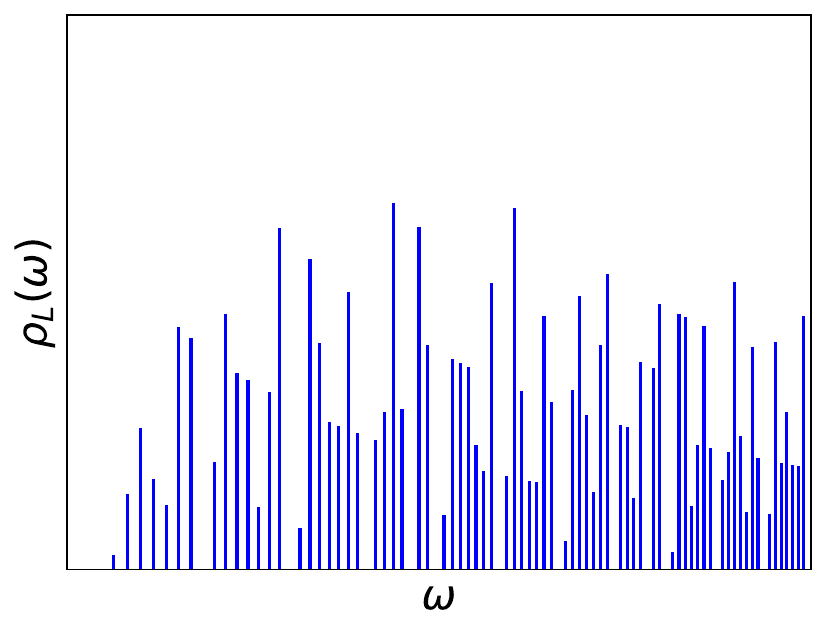}
  \end{subfigure}
  \caption{Sketch of the infinite-volume spectral density $\rho(\omega)$ (left) and the finite-volume $\rho_V(\omega)$ for a specific volume $V$ (right). The height of $\rho_V(\omega)$ represents the multiplicity of the states with the same energy $\omega$.}
  \label{fig:FiniteVolSPectralFunction}
\end{figure}

On the lattice, there is a well-known challenge concerning the reconstruction of the spectral density from correlators $C(t)$ with a finite set of discrete time slices, commonly referred to as the ill-posed inverse problem. Even if the inverse problem could be solved for a correlator in a finite volume, $C_{V}(t)$, where $V=L^3$ denotes the volume of the lattice, and hence the spectral density $\rho_V(\omega)$ is reconstructed, there is still a qualitative difference from its infinite volume counterpart $\rho(\omega)$. The spectral density in the infinite volume is a smooth function, while $\rho_V(\omega)$ is given by a sum of $\delta$-peaks representing allowed states in a finite volume. In Fig. \ref{fig:FiniteVolSPectralFunction} we sketch the situation for two-body states in a finite volume.
This problem is avoided by the introduction of the smearing in the kernel function $K(\omega)$ as shown in Eq. \eqref{equ:KernelFunction}. The inverse problem is made arbitrarily mild by increasing the smearing width $\sigma$, and the smeared spectral density $\rho_{\sigma, V}$ then smoothly approaches its infinite volume counterpart. To recover the inclusive decay rate, we therefore need to take the limit $V\to\infty$ before taking the limit of vanishing smearing width.

The finite-volume effects for the spectral density can be sizeable for multi-body states, because the allowed states are controlled by the boundary condition. The energy spectrum for two-body states, for instance, receives corrections of $\mathcal{O}(1/L^3)$. This would be reduced significantly for the smeared spectral density, but its size and the scaling to the $V\to\infty$ limit may be non-trivial. We therefore introduce a model to investigate the volume dependence. After checking that the model describes the finite-volume data well, we use it to estimate the finite-volume effects.

Among various multi-hadron states, we consider two-body final states, i.e. $K\bar{K}$ states to be specific, which give the dominant contribution. The spectral density is obtained from the imaginary part of the vacuum polarization function, evaluated at one-loop, as
\begin{align}
  \rho(\omega) = \pi \int \frac{d^3\pmb{q}}{(2\pi)^3} \frac{1}{(2\epsilon_{\pmb{q}})^2} \delta(\omega - 2\epsilon_{\pmb{q}})
  \label{equ:IntermediateFormulaSpectrum}
\end{align}
introducing the short-hand notation $\epsilon_{\pmb{q}}^2 = \pmb{q}^2 + m_K^2$. It corresponds to the production of $K\bar{K}$ states from the vacuum through an operator $\mathcal{O}$, which is taken either as a scalar density ($J=0$) or vector current ($J=1$). It models the two-body decays of the $D_s$ meson under an assumption that the wave function of the $D_s$ meson has only insignificant effects, which can be incorporated later by introducing a form factor.

Within this model, one can obtain an explicit expression for the spectral density in the finite-volume and in the infinite-volume limit:
\begin{align}
  \rho_V(\omega) = \frac{\pi}{V} \sum_{\pmb{q}} \frac{1}{4(\pmb{q}^2 + m^2)} \delta\left(\omega - 2\sqrt{\pmb{q}^2 + m^2}\right) \, ,
  \label{equ:SpectralDenScalar}
\end{align}
and
\begin{align}
  \rho(\omega) = \frac{1}{16\pi} \sqrt{1 - \frac{4m^2}{\omega^2}} \, ,
  \label{equ:InfVolSpecScalar}
\end{align}
respectively. The expression above corresponds to the scalar density ($J=0$). For the vector current ($J=1$), we obtain
\begin{align}
  \rho_V(\omega) = \frac{\pi}{V} \sum_{\pmb{q}} \frac{\pmb{q}^2}{4(\pmb{q}^2 + m^2)} \delta\left(\omega - 2\sqrt{\pmb{q}^2 + m^2}\right) \, .
  \label{equ:SpectralDenVec}
\end{align}
and
\begin{align}
  \rho(\omega) = \frac{1}{64\pi} \omega^2 \left(1 - \frac{4m^2}{\omega^2}\right)^{3/2} \, ,
  \label{equ:InfVolSpecVector}
\end{align}

\begin{figure}
  \centering
  \begin{subfigure}{0.49\textwidth}
    \centering
    \includegraphics[width=\textwidth]{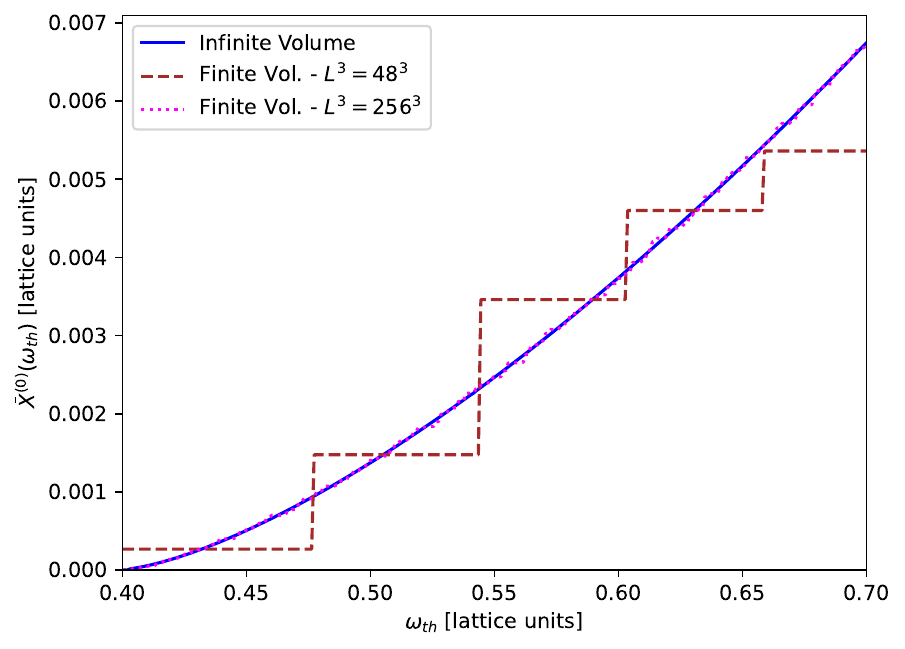}
  \end{subfigure}
  \begin{subfigure}{0.49\textwidth}
    \centering
    \includegraphics[width=\textwidth]{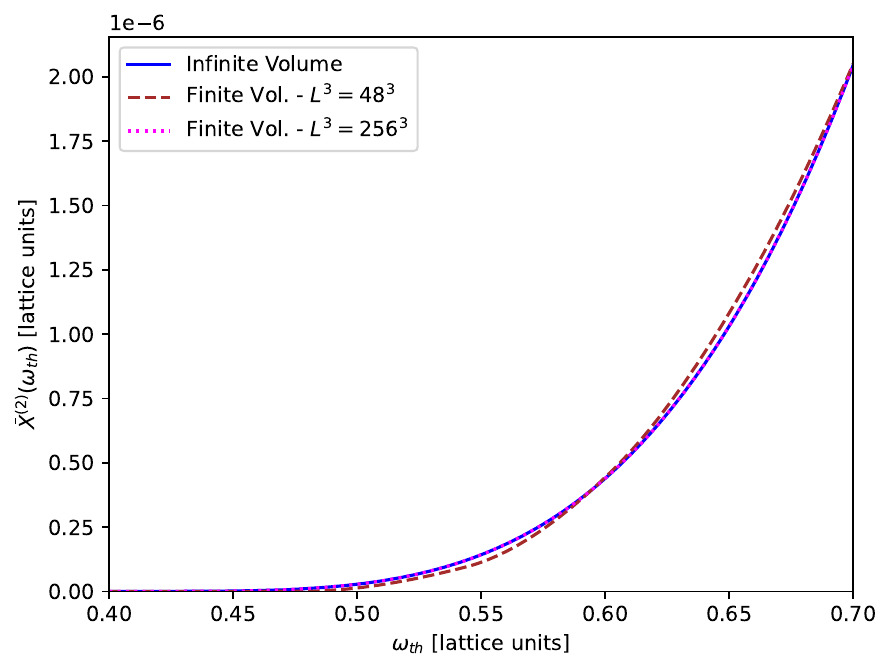}
  \end{subfigure}
  \caption{Infinite volume limit (solid line) of the integral \eqref{equ:AuxilaryFunction} and its finite volume evaluation on $48^3$ (dashed line) and $256^3$ (dash-dotted line) lattices using the finite volume expressions obtained for the spectral density for $J=0$ with $l=0$ (left) and $J=1$ with $l=2$ (right) as a function of the threshold energy $\omega_{\text{th}}$.}
  \label{fig:VolumeReconstruction}
\end{figure}

To estimate how the infinite volume limit is approached, we consider
\begin{align}
  \bar{X}^{(l)}(\omega_{\text{th}}) = \int_0^{\omega_{\text{th}}} d\omega \rho(\omega) \times K^{(l)}(\omega) \, ,
  \label{equ:AuxilaryFunction}
\end{align}
which is defined as a convolution between the kernel and the spectral density. We introduce a variable $\omega_{\text{th}}$, which can be understood as a varying energy cut-off in the kernel function. Although $\omega_{\text{th}}$ is fixed for the physical semileptonic decay process, we use the freedom to choose it in the analysis in order to study how well our model describes the lattice data.
In Fig. \ref{fig:VolumeReconstruction} we show $\bar{X}^{(l)}(\omega_{\text{th}})$ for two choices of the volume $V = 48^3 \text{ and } 256^3$, as well as the infinite volume limit.

We find that the volume effect depends on the choice of $l$ in the kernel function. As argued in \cite{Barone:2023tbl}, due to the sharp cut around the threshold in the kernel for $l=0$ (left), a strong dependence on the volume is expected, while $l=2$ (right) smoothly approaches zero at the threshold and is hence expected to only possesses a mild dependence. Nonetheless, for both cases we observe that $V=256^3$ nearly reproduces the infinite volume expression.

%\section{Lattice Setup}
%\label{sec:Numerics}

%Our simulations employ gauge ensembles from the JLQCD collaboration generated for $2+1$ flavors of dynamical quarks in lattice QCD.
%We perform our simulations on a $64^3 \times 96$ lattice with a lattice spacing of $a \simeq \SI{0.055}{\femto\meter}$ which corresponds to a cut-off of $a \sim \SI{3.610(9)}{\giga\electronvolt}$. We use the M\"obius domain-wall action \cite{Brower:2012vk, Tomii:2016xiv} for both heavy and light quarks. The choice of light quark masses used in this paper corresponds to a pion mass of $M_\pi \simeq \SI{300}{\mega\electronvolt}$. 
%For our simulations, we average over 50 statistically independent gauge configurations and perform our measurement for each configuration at 8 evenly distributed choices of the time source. We induce four different momenta $\pmb{q}$ in the four point function defined. In units of $\pmb{q} = 2\pi/L \pmb{l}$, our induced momenta correspond to $\pmb{l} = (0,0,0), (0,0,1), (0,1,1), (1,1,1)$. All correlation functions analyzed in this work have been simulated using the Grid \cite{Grid:Boyle, Boyle:2015tjk, Yamaguchi:2022feu} and Hadrons \cite{antonin_portelli_2022_6382460} software packages. For most of the fits shown in this paper we have employed lsqfit \cite{lepage:lsqfit, Lepage:2001ym}.

\section{Systematic error due to finite volume effects}
\label{sec:SysErrFinVol}

We combine the model and the lattice data to study the infinite volume limit. We  construct a fit function of the lattice data
\begin{align}
    \bar{C}(t) = A_0 e^{-E_0t} + s(L) \sum_{i} A_i e^{-E_i t} \frac{1}{E_i^2 - m_J^2} \, ,
    \label{equ:ModelFitFunction}
\end{align}
where we treat the ground state separately and sum over the two-body excited states in the second term. The factor $1/(E_i^2 - m_J^2)$ appearing in the second term is motivated by the time-like kaon form factor (pole-dominance model) where the mass $m_J$ is that of the state of corresponding quantum number, i.e. $f_0$ for $J=0$ and $\phi$ for $J=1$.
%Within this fit prescription, all factors except $s(L)$ are under control. 
We constrain the prior of the ground state energy $E_0$ and its amplitude $A_0$ through a fit to the lattice data. The energies and amplitudes $E_i$ and $A_i$ are taken from our model. The prefactor $s(L)$ is determined by a fit to the lattice data, and thus only the relative magnitude of $A_i$'s are relevant.

We consider the case of the spatial current insertions $A_iA_i$ with vanishing $\pmb{q}^2$. This channel contributes only when $l=2$ in the kernel function.

\begin{figure}[t]
	\centering
	\begin{subfigure}{0.49\textwidth}
    	\centering
    	\includegraphics[width=\textwidth]{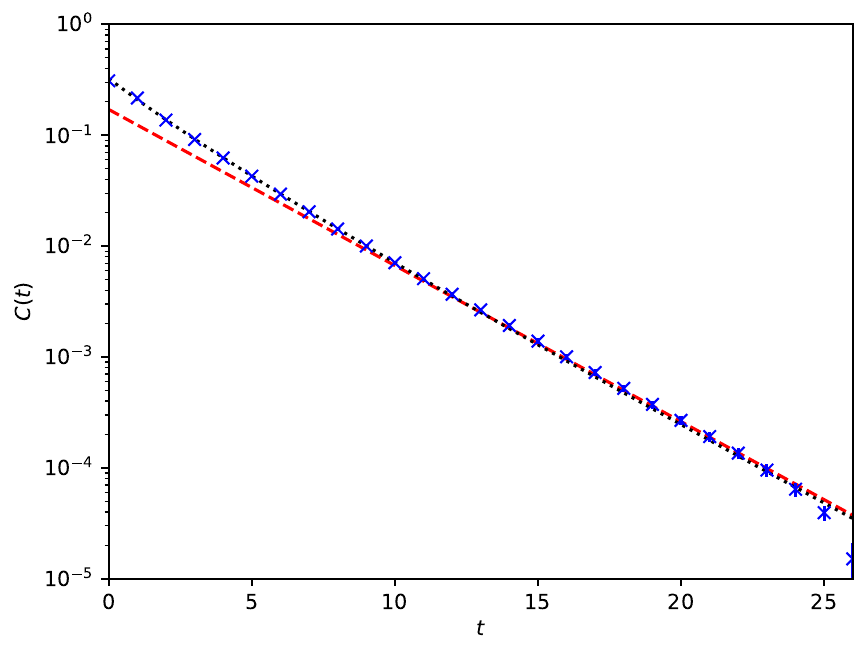}
 	\end{subfigure}
 	\begin{subfigure}{0.49\textwidth}
   		\centering
    	\includegraphics[width=\textwidth]{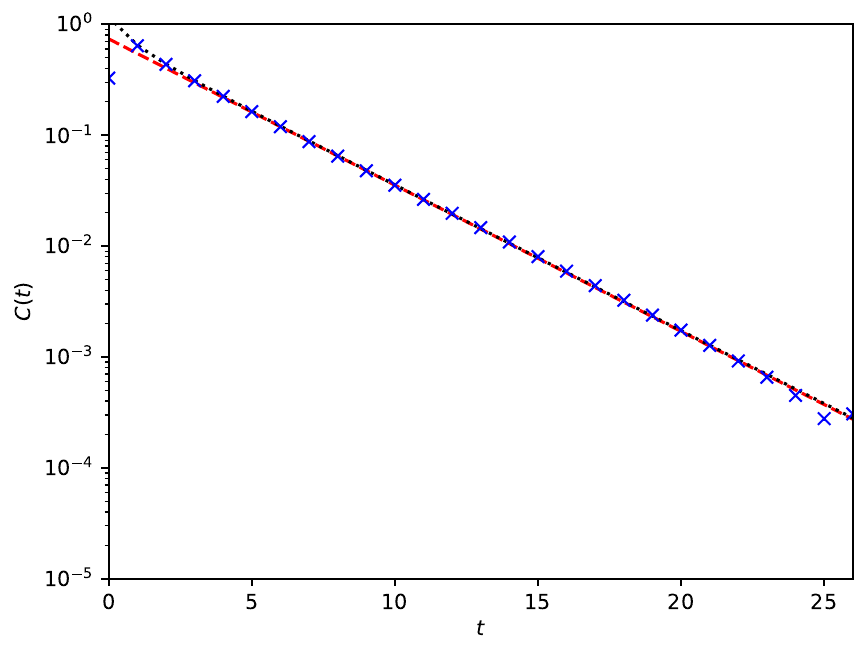}
  	\end{subfigure}
  	\caption{Four-point correlation function for the temporal current insertion of the axial channel (left) and the spatial components (right). For both plots we fit the correlator directly to extract the ground state contribution needed to fix the prior in our model. This fit is represented by the red dashed line. The black dash-dotted line represents the fit results obtained from a fit to our model using the fit prescription \eqref{equ:ModelFitFunction}.}
  	\label{fig:FitToCorrelatorModel}
\end{figure}

In Fig. \ref{fig:FitToCorrelatorModel} we compare the fit results to the lattice data of the four-point correlation function. We also include the fit to the ground state. We observe that the short-distance behavior of the correlator, where the excited states contributions become significant, is well described by our fit, while also reproducing the correct long distance behavior.

\begin{figure}[t]
    \centering
    \begin{subfigure}{0.49\textwidth}
        \includegraphics[width=\textwidth]{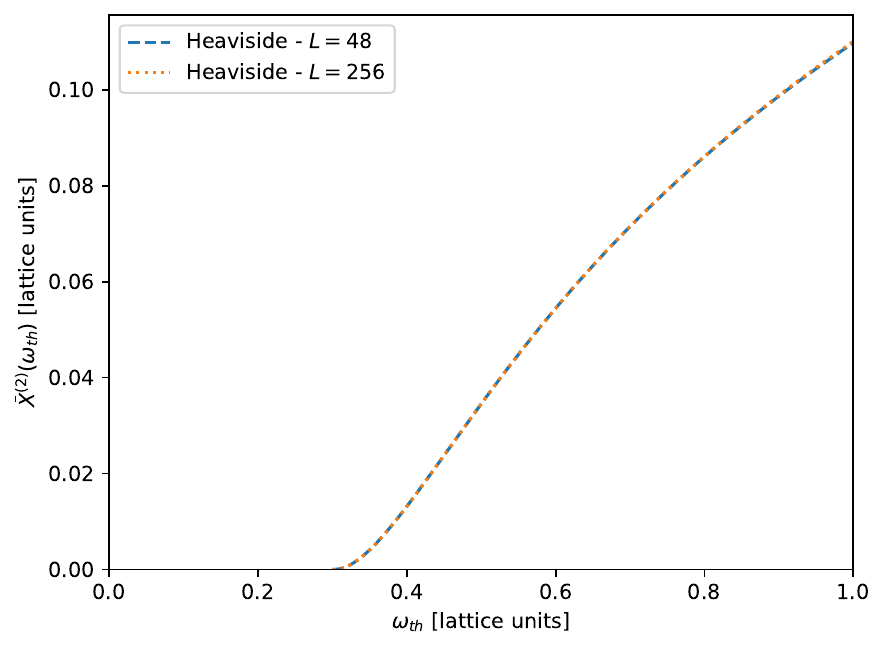}
    \end{subfigure}   
    \begin{subfigure}{0.49\textwidth}
        \includegraphics[width=\textwidth]{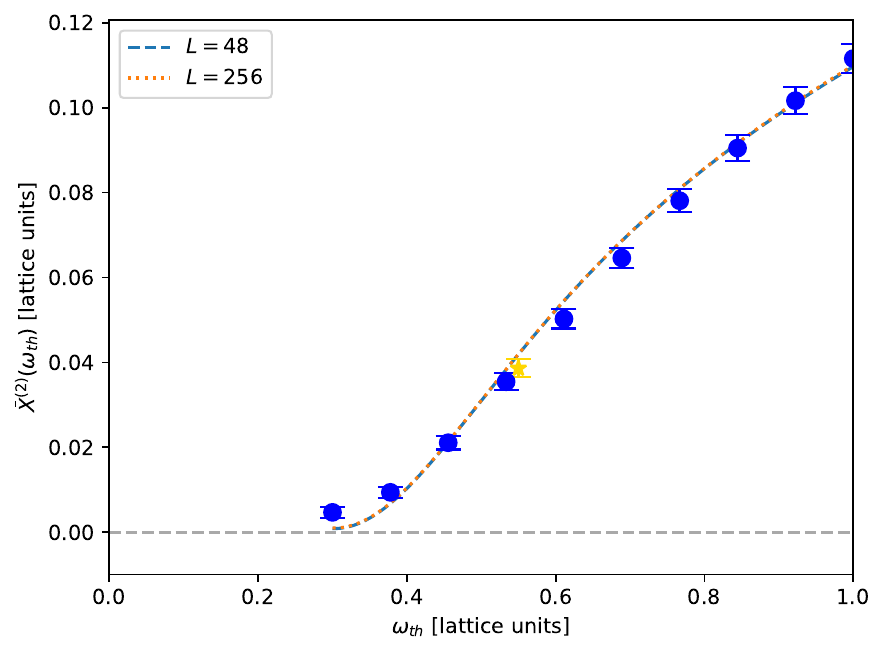}
    \end{subfigure}   
    \caption{Contribution of spatial currents to $\bar{X}_{AA}^{\perp, \parallel}(\pmb{q}^2)$ at $\pmb{q}^2 = \SI{0}{\giga\electronvolt^2}$. We show the results for two choices of the volume $L=48^3$ and $256^3$. The left panel assumes that the cut-off in the kernel function is implemented through a heaviside function, while the left panel assumes the smeared kernel. For the latter, we also compare the results with those obtained from the Chebyshev analysis of our lattice data evaluated for different choices of the threshold $\omega_{\text{th}}$ represented by the blue dots. The physical value of the threshold $\omega_{\text{th}} = \omega_{\text{th}}^{\text{Phys}}$ is denoted by the star symbol.}
    \label{fig:enter-label}
\end{figure}

In Fig. \ref{fig:enter-label} we calculate $\bar{X}^{(l)}(\omega_{\text{th}})$ from \eqref{equ:AuxilaryFunction} using the spectrum determined by the fit. We fit the lattice data at a volume $V=48^3$ and then use it to calculate results for $V=256^3$ which serves as a proxy for the infinite volume limit. On the l.h.s. we combine the fit with the kernel function assuming that the cut is implemented through a Heaviside function. On the r.h.s. we show the case assuming a smeared kernel with a smearing width $\sigma=0.1$. For the latter, we also compare the results with the Chebyshev analysis of the lattice data following \eqref{equ:ChebApprox}, where we repeat the analysis for a set of values of $\omega_{\text{th}}$. We confirm a good agreement between our model and the results obtained from the lattice data.
We conclude that for this specific case the volume dependence is quite mild since no major changes in the shape of the results are found depending on the choice of the volume. 

Finally, we address how we construct our estimate of the corrections to the lattice result. The estimate is constructed by adding the corrections from the $V\to\infty$ limit before adding the $\sigma\to 0$ extrapolation, which translates to: $\bar{X}_{AA}^{\perp}(\pmb{0}^2) = \num{0.0786(31)} \text{ (lattice result) } + 0.0001(0) \text{ (finite volume correction) } + \num{0.0055(1)} \text{ (Finite smearing correction) } = \num{0.0843(31)}$.
%in the following way
%\begin{align}
%	\begin{split}
%		\bar{X}_{AA}^{\perp}(\pmb{0}^2) &= \bar{X}_{A_i A_i}^{(2), \text{Data}}(\pmb{0}^2; \omega_{\text{th}} =\omega_{\text{th}}^{\text{Phys}}, L= 48^3, \sigma = 0.1)  \\
%		&\quad + \underbrace{\bar{X}_{A_i A_i}^{(2), \text{Model}}(\pmb{0}^2; \omega_{\text{th}}^{\text{Phys}}, 256^3, 0.1) - \bar{X}_{A_i A_i}^{(2), \text{Model}}(\pmb{0}^2; \omega_{\text{th}}^{\text{Phys}}; 48^3, 0.1)}_{\text{Finite volume correction}} \\
%		&\quad + \underbrace{\bar{X}_{A_i A_i}^{(2), \text{Model}}(\pmb{0}^2; \omega_{\text{th}}^{\text{Phys}}, 256^3, 0) - \bar{X}_{A_i A_i}^{(2), \text{Model}}(\pmb{0}^2; \omega_{\text{th}}^{\text{Phys}}; 256^3, 0.1)}_{\text{Finite smearing correction}}  \, .
%	\end{split}
%\end{align}
%Translating this into numbers we obtain
%\begin{align}
%	\begin{split}
%		\bar{X}_{AA}^{\perp}(\pmb{0}^2) &= \underbrace{\num{0.0786(31)}}_{\text{Lattice result}} + \underbrace{0.0001(0)}_{\text{Finite volume correction}} + \underbrace{\num{0.0055(1)}}_{\text{Finite smearing correction}} \\
%		&= \num{0.0843(31)}\, .
%	\end{split}
%\end{align}
For the case considered in this work, the corrections due to the infinite volume limit are negligible, while the $\sigma\to 0$ limit gives a correction of order $\sim\SI{7}{\percent}$.

\section{Conclusion}
\label{sec:Conclusion}

We developed a model under the assumption of two-body final states for which we have full control over the infinite volume extrapolation and then combine it with a fit to the lattice data to estimate the expected corrections from the infinite volume limit. In the case study performed here we found negligible corrections due to the infinite volume extrapolation, although larger corrections for larger values of $\pmb{q}^2$ and different shapes of the kernel function are expected. Further work is required to give a proper estimate of the systematic error associated with finite volume effects.

\section*{Acknowledgments}

The numerical calculations of the JLQCD collaboration were performed on SX-Aurora TSUBASA at the High Energy Accelerator Research Organization (KEK) under its Particle, Nuclear and Astrophysics Simulation Program, as well as on Fugaku through the HPCI System Research Project (Project ID: hp220056).

The works of S.H. and T.K. are supported in part by JSPS KAKENHI Grant Numbers 22H00138 and 21H01085, respectively, and by the Post-K and Fugaku supercomputer project through the Joint Institute for Computational Fundamental Science (JICFuS).


\begin{thebibliography}{99}

%\cite{Workman:2022ynf}
%\bibitem{Workman:2022ynf}
%R.~L.~Workman \textit{et al.} [Particle Data Group],
%``Review of Particle Physics,''
%PTEP \textbf{2022} (2022), 083C01
%doi:10.1093/ptep/ptac097

%\cite{Kellermann:2022mms}
\bibitem{Kellermann:2022mms}
R.~Kellermann, A.~Barone, S.~Hashimoto, A.~J\"uttner and T.~Kaneko,
%``Inclusive semi-leptonic decays of charmed mesons with M\"obius domain wall fermions,''
PoS \textbf{LATTICE2022}, 414 (2023)
doi:10.22323/1.430.0414
[arXiv:2211.16830 [hep-lat]].
%4 citations counted in INSPIRE as of 15 Nov 2023

%\cite{Gambino:2020crt}
\bibitem{Gambino:2020crt}
P.~Gambino and S.~Hashimoto,
%``Inclusive Semileptonic Decays from Lattice QCD,''
Phys. Rev. Lett. \textbf{125} (2020) no.3, 032001
doi:10.1103/PhysRevLett.125.032001
[arXiv:2005.13730 [hep-lat]].

%\cite{Gambino:2022dvu}
\bibitem{Gambino:2022dvu}
P.~Gambino, S.~Hashimoto, S.~M\"achler, M.~Panero, F.~Sanfilippo, S.~Simula, A.~Smecca and N.~Tantalo,
%``Lattice QCD study of inclusive semileptonic decays of heavy mesons,''
JHEP \textbf{07} (2022), 083
doi:10.1007/JHEP07(2022)083
[arXiv:2203.11762 [hep-lat]].

%\cite{Hansen:2017mnd}
\bibitem{Hansen:2017mnd}
M.~T.~Hansen, H.~B.~Meyer and D.~Robaina,
%``From deep inelastic scattering to heavy-flavor semileptonic decays: Total rates into multihadron final states from lattice QCD,''
Phys. Rev. D \textbf{96} (2017) no.9, 094513
doi:10.1103/PhysRevD.96.094513
[arXiv:1704.08993 [hep-lat]].
%75 citations counted in INSPIRE as of 30 Nov 2022

%\cite{Hansen:2019idp}
\bibitem{Hansen:2019idp}
M.~Hansen, A.~Lupo and N.~Tantalo,
%``Extraction of spectral densities from lattice correlators,''
Phys. Rev. D \textbf{99} (2019) no.9, 094508
doi:10.1103/PhysRevD.99.094508
[arXiv:1903.06476 [hep-lat]].
%22 citations counted in INSPIRE as of 21 Oct 2022

%\cite{Bulava:2021fre}
\bibitem{Bulava:2021fre}
J.~Bulava, M.~T.~Hansen, M.~W.~Hansen, A.~Patella and N.~Tantalo,
%``Inclusive rates from smeared spectral densities in the two-dimensional O(3) non-linear \ensuremath{\sigma}-model,''
JHEP \textbf{07} (2022), 034
doi:10.1007/JHEP07(2022)034
[arXiv:2111.12774 [hep-lat]].
%8 citations counted in INSPIRE as of 21 Oct 2022

%\cite{Barone:2023tbl}
\bibitem{Barone:2023tbl}
A.~Barone, S.~Hashimoto, A.~J\"uttner, T.~Kaneko and R.~Kellermann,
%``Approaches to inclusive semileptonic B$_{(s)}$-meson decays from Lattice QCD,''
JHEP \textbf{07}, 145 (2023)
doi:10.1007/JHEP07(2023)145
[arXiv:2305.14092 [hep-lat]].
%4 citations counted in INSPIRE as of 15 Nov 2023

%\cite{Bulava:2023mjc}
%\bibitem{Bulava:2023mjc}
%J.~Bulava,
%``The spectral reconstruction of inclusive rates,''
%PoS \textbf{LATTICE2022}, 231 (2023)
%doi:10.22323/1.430.0231
%[arXiv:2301.04072 [hep-lat]].
%6 citations counted in INSPIRE as of 20 Nov 2023

%\cite{Brower:2012vk}
%\bibitem{Brower:2012vk}
%R.~C.~Brower, H.~Neff and K.~Orginos,
%``The M\"obius domain wall fermion algorithm,''
%Comput. Phys. Commun. \textbf{220}, 1-19 (2017)
%doi:10.1016/j.cpc.2017.01.024
%[arXiv:1206.5214 [hep-lat]].
%150 citations counted in INSPIRE as of 16 Nov 2023

%\cite{Tomii:2016xiv}
%\bibitem{Tomii:2016xiv}
%M.~Tomii \textit{et al.} [JLQCD],
%``Renormalization of domain-wall bilinear operators with short-distance current correlators,''
%Phys. Rev. D \textbf{94}, no.5, 054504 (2016)
%doi:10.1103/PhysRevD.94.054504
%[arXiv:1604.08702 [hep-lat]].
%29 citations counted in INSPIRE as of 16 Nov 2023

%\bibitem{Grid:Boyle}
%Peter Boyle, Azusa Yamaguchi, Guido Cossu, and Antonin Portelli. Grid: Data parallel C++ mathematical object library. https://github.com/paboyle/Grid.

%\cite{Boyle:2015tjk}
%\bibitem{Boyle:2015tjk}
%P.~Boyle, A.~Yamaguchi, G.~Cossu and A.~Portelli,
%``Grid: A next generation data parallel C++ QCD library,''
%[arXiv:1512.03487 [hep-lat]].
%57 citations counted in INSPIRE as of 16 Nov 2023

%\cite{Yamaguchi:2022feu}
%\bibitem{Yamaguchi:2022feu}
%A.~Yamaguchi, P.~Boyle, G.~Cossu, G.~Filaci, C.~Lehner and A.~Portelli,
%``Grid: OneCode and FourAPIs,''
%PoS \textbf{LATTICE2021}, 035 (2022)
%doi:10.22323/1.396.0035
%[arXiv:2203.06777 [hep-lat]].
%10 citations counted in INSPIRE as of 16 Nov 2023

%\bibitem{antonin_portelli_2022_6382460}
%Antonin Portelli, Ryan Abott, Nils Asmussen, et al. Hadrons: Grid-based workflow man- agement system for lattice field theory simulations. https://github.com/aportelli/Hadrons.

%\bibitem{lepage:lsqfit}
%Peter Lepage and Christoph Gohlke. gplepage/lsqfit: lsqfit version 12.0.3, December 2021.

%\cite{Lepage:2001ym}
%\bibitem{Lepage:2001ym}
%G.~P.~Lepage \textit{et al.} [HPQCD],
%``Constrained curve fitting,''
%Nucl. Phys. B Proc. Suppl. \textbf{106}, 12-20 (2002)
%doi:10.1016/S0920-5632(01)01638-3
%[arXiv:hep-lat/0110175 [hep-lat]].
%312 citations counted in INSPIRE as of 16 Nov 2023

\end{thebibliography}
\end{document}